\documentclass[12pt,pra]{revtex4}
\usepackage{bm,amsfonts,amssymb, graphicx}
\begin{document}
\title{Rapid refractive index enhancements via laser-mediated collectivity}

\author{Mihai~\surname{Macovei}\footnote{Permanent address:\it{Technical University of Moldova, 
Physics Department, \c{S}tefan Cel Mare Av. 168, MD-2004 Chi\c{s}in\u{a}u, Moldova.}}}
\email{macovei@usm.md}

\author{Christoph~H.~\surname{Keitel}}
\email{keitel@mpi-hd.mpg.de}

\affiliation{Max-Planck-Institut f\"ur Kernphysik, 
 Saupfercheckweg 1, D-69117 Heidelberg, Germany}

\date{\today}
\begin{abstract}
The collective interaction via the environmental vacuum is investigated 
for a mixture of two deviating multi-atom ensembles in a moderately intense 
laser field. Due to the numerous inter-atomic couplings, the laser-dressed 
system may react sensitively and rapidly with respect to changes in 
the atomic and laser parameters. We show for weak probe fields that in the 
absence of absorption both the index of refraction and the group velocity 
may be modified strongly and rapidly due to the collectivity.
\end{abstract}
\maketitle

The presence of strong laser fields are known to substantially modify the absorptive and dispersive properties of atomic samples \cite{Moll,Boyd,H_EIT,LWI,refr1,refr2,stop,ultra,HarPl,AgL,Man,mek}. The numerous effects put forward already for single atoms include the splitting of spectral lines \cite{Moll,Boyd}, electromagnetically induced transparency \cite{H_EIT}, lasing without population inversion \cite{LWI}, enhanced indeces of refraction \cite{refr1,refr2}, stopping of light \cite{stop} and ultra-narrow spectral lines \cite{ultra}. For larger ensembles and not too low densities, collective phenomena were further pointed out to drastically affect laser - driven media \cite{HarPl,AgL,mek,Andr,Puri}. In particular, collectivity due to nonlinearities in a plasma may render the medium transparent \cite{HarPl}, while laser-mediated local effects arising from dipole-dipole interactions between atoms may alter the appearance of band gaps in optically dense materials \cite{AgL} and induce piezophotonic switching \cite{Man}. Efficient schemes to control the collective quantum dynamics in general were further demonstrated \cite{mek} via employing interferences and fast switching schemes among collective atomic dressed-states. Particular interest, however, is in the rapid control of dispersive properties of collective systems, e.g. for quantum gates or high precision measurements \cite{Dan}. 

In this letter we show that the mutual interactions of atoms via the quantum fluctuations of the surrounding electromagnetic field (EMF) \cite{Dicke} are suitable to generate transparent media with large indeces of refraction of order 10 or high dispersion of arbitrary sign.  In particular we point out that the effects may be considerably larger than for the same amount of independent atoms and may be set up on a time scale of 10$^{-9}$s (GHz). Furthermore, the group velocity of a weak probe field propagating through the strongly driven two-level medium may be substantially slowed down or strongly accelerated. 

\begin{figure}[b]
\begin{center}
\includegraphics[width=7.7cm]{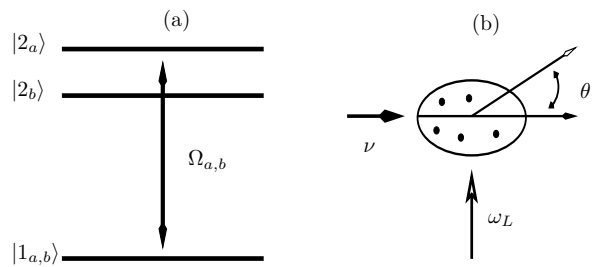}
\end{center}
\caption{\label{fig-1} Schematic diagram depicting means of control over refractive properties of two laser-driven two-level atomic ensembles. (a) The involved energy levels of the two two-level atomic systems and Rabi frequencies are denoted with $a$ and $b$. (b) The index of refraction and thus the deviation angle $\theta$ of a weak probe field ($\nu$) may be substantially and rapidly manipulated by an applied laser field ($\omega_{L}$).}
\end{figure}

For this purpose, we consider an atomic system consisting of two ensembles of two-level atoms, numbered $N_{a}$ and $N_{b}$, with densities of order $10^{12}-10^{14}cm^{-3}$, and with somewhat different transition frequencies, and interacting with a single moderately strong laser field (see Fig.~\ref{fig-1}). The corresponding Rabi frequencies are $\{ 2\Omega_{a}, 2\Omega_{b}\}$ and spontaneous decay of all closely spaced atoms occurs via interaction with a common electromagnetic field reservoir with rates $\{2\gamma_{a}, 2\gamma_{b}\}$ from excited states $\{|2_{a} \rangle, |2_{b}\rangle\}$, respectively. In order to treat the atoms uniformly we suppose that $\{L/c, \Omega^{-1}_{a,b} \ll \tau_{s}\},$ where $L$, $c$ and $\tau_{s}$ are the largest dimension of the sample, the light velocity and the collective decay time, respectively. When scanning the composed atomic sample with, say for instance, a dye or diode laser \cite{Lexp}, and depending on the resonance condition for each kind of atom and thus the employed laser frequency, generally one or the other atom specy may dominate the final steady-state collective behavior. 

In the conventional mean-field and rotating-wave approximation, the interaction of the laser-driven atomic sample with the surrounding EMF bath is described in a frame rotating with the laser frequency $\omega_{L}$, via the Hamiltonian: 
\begin{eqnarray}
H = H_{f} + H_{0} + H_{i}, \label{Hm}
\end{eqnarray}
where $H_{f}=\sum_{k}\hbar(\omega_{k}-\omega_{L})a^{\dagger}_{k}a_{k}$, $H_{0}$ = $\sum_{j}\hbar [\Delta_{j}S^{(j)}_{z}+\Omega_{j}(S^{(j)}_{+}+S^{(j)}_{-})]$ and 
$H_{i}$=$i\sum_{k}\sum_{j}(\vec g_{k}\cdot \vec d_{j})\bigl (a^{\dagger}_{k}S^{(j)}_{-} - a_{k}S^{(j)}_{+} \bigr).$ Here the first and the second terms in Eq.~(\ref{Hm}) represent the free EMF ($H_{f}$) and the free atomic plus laser-atom interaction Hamiltonian ($H_{0}$), respectively. The last term of Eq.~(\ref{Hm}) describes the interaction ($H_{i}$) of the atoms with the environmental vacuum modes ($a_k$). The collective atomic operators $S^{(i)}_{+} = \sum^{N_{i}}_{l=1}|2_{i}\rangle_{l}{}_{l}\langle 1_{i}|$ $\equiv |2_{i}\rangle \langle 1_{i}|,$ $S^{(i)}_{-}$ = $[S^{(i)}_{+}]^{\dagger}$, $S^{(i)}_{z}$ = $\frac{1}{2}\sum^{N_{i}}_{l=1}(|2_{i}\rangle_{l}{}_{l}\langle 2_{i}| - |1_{i}\rangle_{l}{}_{l}\langle 1_{i}|)$ $ \equiv \frac{1}{2}(|2_{i}\rangle \langle 2_{i}| - |1_{i}\rangle \langle 1_{i}|)$ satisfy the standard commutation relations for quasispin operators, i.e., $[S^{(i)}_{z},S^{(j)}_{\pm}]$ = $\pm \delta_{ij}S^{(i)}_{\pm}$,  $[S^{(i)}_{+},S^{(j)}_{-}]$ = 2$\delta_{ij}S^{(i)}_{z}$, $(\{i,j\} \in \{a,b \})$. $\Delta_{i}$ = $\omega_{i} - \omega_{L}$ denote the detuning of the atomic transitions frequencies $\omega_{i}$ to the laser frequency, and $\{\Delta_{i},\Omega_{i} \ll \omega_{i}\}$. $d_{i}$ corresponds to the transition dipole matrix elements of the atoms. $\vec g_{k}=\sqrt{2\pi\hbar\omega_{k}/V}\vec e_{\lambda},$ where $\vec e_{\lambda}$ is the photon polarization vector while $V$ is the EMF quantization volume.  

The master equation corresponding to the Hamiltonian (\ref{Hm}) in the Born-Markov approximation then reads
\begin{eqnarray}
\dot \rho(t) &+& \frac{i}{\hbar}[H_{0},\rho] = -\sum_{i,j \in \{a,b\}}\bigl (\sqrt{\gamma_{i}\gamma_{j}}[S^{(i)}_{+},S^{(j)}_{-}\rho] \nonumber \\
&+& \sqrt{r_{i}r_{j}}[S^{(i)}_{z},S^{(j)}_{z}\rho] \bigr )+ h.c. \label{Meq}
\end{eqnarray}
The diagonal ($i=j$) contribution of the first term on the right-hand side describes the collective damping due to the spontaneous emission of atoms, while that proportional to $\sqrt{\gamma_{a}\gamma_{b}}$ involves the mutual exchange of photons among the different types of atoms in the sample and is very sensitive relative to the splitting frequency $\Delta \omega$ = $\omega_{a}$ - $\omega_{b}$ with $\Delta_{a}-\Delta_{b} = \Delta \omega$. The dipole-dipole interactions $\Theta_{dd}$ among the emitters are omitted here, an approximation valid as soon as $\Omega_{a,b}/N_{a,b} \gg \Theta_{dd}$. The last term of the master equation Eq.~(\ref{Meq}) involving the collision rates ${r_{i}}$ accounts for collisional damping of atoms which alter the phase of the atomic state but not its population \cite{Puri,SZb}.

In the intense-field limit ($\Omega_{i} \gg \gamma_{i}N_{i}$), the master equation  Eq.~(\ref{Meq}) transforms into the dressed-state picture ($|\Psi^{(i)}_{j}\rangle$
for $\{i \in a,b \}$, $\{j \in 1,2 \}$) via
\begin{eqnarray}
|1_{i}\rangle &=& |\Psi^{(i)}_{1}\rangle \cos{\theta_{i}} +  |\Psi^{(i)}_{2}\rangle \sin{\theta_{i}}, \nonumber \\ 
|2_{i}\rangle &=& -|\Psi^{(i)}_{1} \rangle \sin{\theta_{i}} +  |\Psi^{(i)}_{2} \rangle \cos{\theta_{i}}, \label{drS} 
\end{eqnarray}
with $\cot{2\theta_{i}}$ = $\Delta_{i}/2\Omega_{i}$, $\{i \in a,b \}$.  In the secular approximation, i.e. upon omission of the terms oscillating with Rabi frequency $\tilde \Omega_{a(b)} = \sqrt{\Omega^{2}_{a(b)} + (\Delta_{a(b)}/2)^{2}}$ and larger, and without the cross-damping contribution ($\gamma_{ab} = \sqrt{\gamma_{a}\gamma_{b}} = r_{ab} = \sqrt{r_{a}r_{b}}=0$), the resulting dressed state master equation results in the exact steady-state solution of the form 
\begin{eqnarray}
\rho_{s} = Z^{-1}\prod_{i \in \{a,b\}}e^{-\xi_{i}R^{(i)}_{z}}. \label{Msol}
\end{eqnarray}
Here $2\xi_{i} = \ln\bigl([\gamma_{i}\cos^{4}{\theta_{i}}+r_{i}\sin^{2}(2\theta_{i})/4]/[\gamma_{i}\sin^{4}{\theta_{i}}+r_{i}\sin^{2}(2\theta_{i})/4] \bigr)$, $R^{(i)}_{z}$ = $|\Psi^{(i)}_{2}\rangle \langle \Psi^{(i)}_{2}|$ - $|\Psi^{(i)}_{1}\rangle \langle \Psi^{(i)}_{1}|$, $\{i \in a,b\},$ while $Z$ is chosen such that ${\rm Tr}{\rho_{s}}=1.$

The influence of the cross-damping terms with respect to the final collective steady-state dynamics can be estimated approximately for larger samples, while for single - atom systems ($N_{a}= N_{b}= 1$) this can be carried out exactly by solving the respective equations of motion for the atomic variables.  In the dressed master equation, the sideband contribution proportional to $ \gamma_{ab}, r_{ab}$ oscillates at the relative frequency $2|\tilde \Omega_{a} - \tilde \Omega_{b}|$ which can be nonzero as, in general, the atoms are subject to different detunings and laser intensities. If $|\tilde \Omega_{a} - \tilde \Omega_{b}| \sim \Omega_{a,b}$ such contributions can be omitted in the secular approximation (as well as for smaller samples), and the solution in Eq. (\ref{Msol}) is then applicable. When $\Omega_{a} \approx \Omega_{b}$ and the strong laser field being detuned far off the frequency range $0 \le \Delta \le \Delta \omega$, i.e. $|\tilde \Omega_{a} - \tilde \Omega_{b}| \approx 0$, then Eq.~(\ref{Msol}) can be employed with $\xi_{a} \approx \xi_{b}$. Assuming $\Delta_{a} = |\Delta_{b}| =: \Delta \omega /2$, and $\Omega_{a} = \Omega_{b} \gg \Delta \omega/2$, the steady-state solution can further be obtained from Eq.~(\ref{Msol}) in the limit $\xi_{a,b} \to 0$. In what follows, however, we neglect the cross-damping contributions and thus restrict ourselves to the case  $N\gamma_{a,b} \le \Delta \omega < \tilde \Omega_{a,b}$.

On employing the atomic coherent states for two-level atoms \cite{ACS} and the solution in Eq.~(\ref{Msol}), we derive the expectation values for any collective atomic correlators of interest. In particular, the steady-state expectation values for the collective dressed-state inversion operators $\langle R^{(i)}_{z}\rangle_{s}$ can be obtained for $\{i \in a,b\}$ and
with $Z = Z_{a}Z_{b}$: 
\begin{eqnarray}
\langle R^{(i)}_{z}\rangle_{s} = - \frac{\partial}{\partial \xi_{i}} \ln Z_{i}, ~~~~ Z_{i}=e^{\xi_{i}N_{i}}\frac{1-e^{-2\xi_{i}(N_{i}+1)}}{1-e^{-2\xi_{i}}}. \nonumber
\end{eqnarray}

In what follows we concentrate on the case with almost equal parameters for the two types of atoms, i.e. $\Omega_{a} \approx \Omega_{b} \equiv \Omega$, $\gamma_{b} \approx \gamma_{a}(1-\Delta\omega/\omega_{a})^{3} \approx \gamma_{a} \equiv \gamma$, $r_{a} \approx r_{b} \equiv r$ and $N_{a} \approx N_{b} \equiv N$. 
\begin{figure}[t]
\begin{center}
\includegraphics[height=3.2cm,width = 8.6cm]{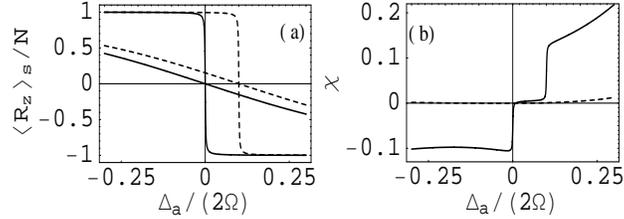}
\caption{\label{fig-2ab} (a) The steady-state dependence of the dressed-state inversion operator $\langle R_{z}\rangle_{s}/N$ versus $\Delta_{a}/(2\Omega)$. The solid and dashed lines stand for atoms of type $a$ and $b$, respectively. Here $r/\gamma =0.3$, $\Delta \omega/(2\Omega)=0.1$. The exterior curves correspond to $N_{a}= N_{b}= 1000$ and the interior to $N_{a}= N_{b} = 1$. (b) The susceptibility $\chi = \chi^{'} + i\chi^{''}$ [in units of $\bar N d^{2}/\gamma \hbar$] as a function of $\Delta_{a}/(2\Omega).$  The solid curve represents $\chi^{'}$ while the dashed $\chi^{''}.$ The remaining parameters are $2\Omega/(N\gamma)=10$, $N_{a} = N_{b} = 1000$ and $\nu-\omega_{a}=0.35\times 2\Omega.$}
\end{center}
\end{figure}

Fig.(\ref{fig-2ab}) depicts the steady-state dependence of the dressed-state inversion operators as a function of the ratio $\Delta_{a}/(2\Omega)$. As the atomic transition frequencies for the atoms of type $a$ and $b$ differ from each other, the steady-state collective populations behave differently as well. Note that, due to collectivity, the collective dressed state populations may be transferred abruptly and rapidly from one dressed state to another as the laser detuning $\Delta_{i}$ changes its sign. A few- atom system is less sensitive relative to the laser detuning in the sense of fast switching. Thus, at this particular point,  $\Delta_{i}/(2\Omega_{i})= \pm \varepsilon$ with $\varepsilon \ll 1,$ one may switch the absorption properties of a weak probe field from positive to negative gain (or vice versa) while the dispersive features are strongly enhanced. For a pencil-shaped sample with length $L \sim 5\lambda$, transversal area $S \sim 2\lambda^{2}$, $\lambda \sim 10^{-4}cm$, $\gamma \sim 10^{7}Hz$, and $N \sim 10^{3}$ we estimate a switching time $\tau_{s} \sim 2L/(\lambda \gamma N)$ of about $10^{-9}s$. The secular approximation can be applied here if $\Omega \sim 10^{10}Hz$, as $\Omega \gg \tau^{-1}_{s}.$

We proceed by calculating the refractive properties of a very weak field probing the strongly driven atomic sample. The linear susceptibility $\chi(\nu)$ of the probe field, at frequency $\nu$, can be represented in terms of the Fourier transform of the average value of the two-time commutator of the atomic operator as
\begin{eqnarray}
\chi(\nu)=\frac{i}{\hbar}\sum_{j \in \{a,b\}}\frac{d^{2}_{j}}{V_{j}}\int^{\infty}_{0}d\tau e^{i\nu \tau}\langle[S^{(j)}_{-}(\tau), S^{(j)}_{+}]\rangle_{s}. \label{chi}
\end{eqnarray}
Note that the steady-state (subindex $s$) of the atomic correlators in Eq.~(\ref{chi}) should be calculated with the help of Eqs.~(\ref{drS},\ref{Msol}).

Introducing Eq.~(\ref{drS}) in Eq.~(\ref{chi}), and making use of both the secular approximation and the quantum regression theorem \cite{SZb}, together with Eq.~(\ref{Msol}), then the dispersion and absorption features can be described via:
\begin{eqnarray}
\chi^{'}(\Delta_{p}) &=& \sum_{i \in \{a,b\}}\frac{\bar N_{i}d^{2}_{i}}{\gamma_{i}\hbar}\frac{\langle R^{(i)}_{z}\rangle_{s}}{N_{i}} \bigl [\cos^{4}{\theta_{i}}\frac{\tilde \Delta^{(i)}_{p} - 2\bar \Omega_{i}}{\tilde \gamma^{2}_{i} + (\tilde \Delta^{(i)}_{p}-2\bar \Omega_{i})^{2}} \nonumber \\
&-&\sin^{4}{\theta_{i}}\frac{\tilde \Delta^{(i)}_{p} + 2\bar \Omega_{i}}{\tilde \gamma^{2}_{i} + (\tilde \Delta^{(i)}_{p} + 2\bar \Omega_{i})^{2}} \bigr], \nonumber \\
\chi^{''}(\Delta_{p}) &=& \sum_{i \in \{a,b\}}\frac{\bar N_{i}d^{2}_{i}}{\gamma_{i}\hbar}\frac{\langle R^{(i)}_{z}\rangle_{s}}{N_{i}} \bigl [\sin^{4}{\theta_{i}}\frac{\tilde \gamma_{i}}{\tilde \gamma^{2}_{i}+(\tilde \Delta^{(i)}_{p} + 2\bar \Omega_{i})^{2}} \nonumber \\
&-&\cos^{4}{\theta_{i}}\frac{\tilde \gamma_{i}}{\tilde \gamma^{2}_{i}+(\tilde \Delta^{(i)}_{p} - 2\bar \Omega_{i})^{2}} \bigr]. \label{refr}
\end{eqnarray} 
Here $\bar \Omega_{i}=\tilde \Omega_{i}/(\gamma_{i}N_{i})$, while $\bar N_{i}$ is the atomic density. $\tilde \Delta^{(i)}_{p}=\Delta_{p}/(\gamma_{i}N_{i})=(\nu - \omega_{L})/(\gamma_{i}N_{i})$ = $(\nu - \omega_{a} + \Delta_{a})/(\gamma_{i}N_{i})$ corresponds to the detuning of the weak probe field frequency with respect to the driving field, while $\tilde \gamma_{i} = (\gamma^{(i)}_{s} - \gamma^{(i)}_{c})/(\gamma_{i}N_{i})$ describes the non-diagonal collective damping with $\gamma^{(i)}_{s}=\gamma_{i}[\sin^{2}(2\theta_{i})+\cos^{4}{\theta_{i}}+\sin^{4}{\theta_{i}}]$ + $r_{i}[\cos^{2}(2\theta_{i})+\sin^{2}(2\theta_{i})/2]$ and $\gamma^{(i)}_{c} =  \gamma_{i}\cos(2\theta_{i})\langle R^{(i)}_{z}\rangle_{s},$ respectively. It should further  be noted that in Eqs.~(\ref{refr}) we have employed the so-called decoupling scheme for symmetrical atomic correlators as valid for $N \gg 1$ \cite{Andr}. In the absence of collective effects ($\gamma^{(i)}_{c} \equiv 0$), Eqs.~(\ref{refr}) reduces, as it should, to the correct result for $N_{a}$ and $N_{b}$ independent atoms. 

\begin{figure}[t]
\begin{center}
\includegraphics[width=8.4cm,height=6cm]{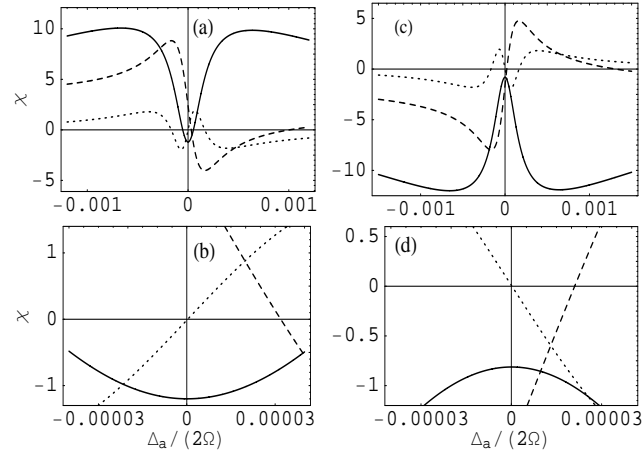}
\end{center}
\caption{\label{fig-3ab} The steady-state dependence of the linear susceptibility $\chi$ (in units of $\bar N d^{2}/\gamma \hbar$ ) as well as that of the derivation $(d/d\nu)\chi^{'}$ (in units of $\bar N d^{2}/\gamma^{2}\hbar$ ) as a function of $\Delta_{a}/2\Omega$. The solid and dashed lines correspond to the real and imaginary parts of $\chi$, respectively, while the dotted curve stands for $(d/d\nu)\chi^{'}$. Here $\nu -\omega_{a}= -2\Omega$ (a,b),  and $\nu -\omega_{a}= 2\Omega$ (c,d), while $N = 1000$, $2\Omega/(N\gamma)=10$, $\Delta \omega/(2\Omega)=0.1$, and $r/\gamma = 0.3$. (b,d) are enlargements of (a,c).}
\end{figure}

On inspecting Eqs.~(\ref{refr}) (involving $\langle R^{(i)}_{z}\rangle_{s}$) and Fig.~(\ref{fig-2ab}a) one can easily recognize that the susceptibility $\{\chi^{'},\chi^{''}\}$ is substantially enhanced via collective effects. Figures (\ref{fig-3ab}) depict the steady-state dependence of the linear susceptibility with respect to the strong laser detunings while keeping fixed probe-field frequencies. Strong gain, strong positive or negative dispersion with zero absorption are then feasible. 
The interpretation of these results is straight forward via a dressed-state analysis \cite{Moll,Boyd}. When $\nu - \omega_{a} = -2\Omega$ (see Fig.~\ref{fig-3ab}a,b), the probe field at exact resonance with the dressed-state transition $|\Psi^{(a)}_{1}\rangle \leftrightarrow |\Psi^{(a)}_{2}\rangle$. If $\Delta_{a}/(2\Omega) <0$ most population is placed in the dressed - states $|\Psi^{(i)}_{2}\rangle$ (see Fig.~\ref{fig-2ab}a) and, thus, the probe field is absorbed. 
Here $\Delta_{a}/(2\Omega)=0$ means that $\langle R^{(a)}_{z}\rangle_{s}=0$ while $\langle R^{(b)}_{z}\rangle_{s} \not =0$ and, respectively, the susceptibility is small though nonzero. Note that for a single-type atomic ensemble one can achieve complete transparency at this point. Increasing further $\Delta_{a}$, i.e. $\Delta_{a}/(2\Omega) >0$, the dressed -state population is transferred completely to $|\Psi^{(a)}_{1}\rangle$ and the probe field is amplified. Thus, the second ensemble contributes here to a strong shift of the susceptibility resulting in zero absorption with large dispersive features. In particular, the index of refraction yield with close to vanishing absorption $n(\nu) \approx \sqrt{1 + \chi^{'}(\nu)}$ \cite{SZb} which for the sample parameters given above takes values larger than $n > 8$ [see Fig.~(\ref{fig-3ab}a) near $\Delta_{a}/(2\Omega)\approx 10^{-3}$].  However, without collective effects, i.e. a noninteracting ensemble of $N_{a}$ = $N_{b}=1000$ atoms, the index of refraction would only be close to unity at the point of vanishing absorption. The index of refraction $n$ can be further enhanced by increasing $\lambda^{3}\bar N$, but then the atoms would be such close to each other that short-range dipole-dipole interactions need be taken into account. Moreover, $\chi^{'}$ may be smaller than  $- 1$ at frequencies where the absorption vanishes meaning that the weak-probe field can not propagate anymore through the atomic medium (see Fig.~\ref{fig-3ab}). 

Furthermore, once the probe field frequency is adjusted near resonance with the bare-state transition frequencies of atoms $a$ or $b$, one can observe steep dispersive features at $\Delta_{a} \in \{ 0,\Delta \omega \}$ (see Fig.~\ref{fig-2ab}b). However, the magnitude of the susceptibility, in this case, is the same as for a few-atom sample, except the dispersive slopes. These results can be employed to induce  a rapid phase shift for a weak field travelling through the strongly driven atomic sample. Note that at the points, where the $\Delta_{i}$ change the sign, the two-level emitters are in a strong collective phase \cite{Puri} and, thus, abrupt changes of $\chi^{'}$ are due to strong collectivity. Also, the collisional damping does not affect considerably the collective steady-state behavior, and its influence can be balanced by increasing the number of atoms. This means that the collisional damping influences small atomic samples while larger atomic systems are less sensitive with regard to kind of phase damping.   

We demonstrate further that collections of two-level atoms are suitable for rapidly switching between strongly accelerating or slowing down of a weak probe pulse traversing through driven two-level media. The light group velocity can be estimated from the following expressions: $1/v_{g}=n_{g}/c=dk(\nu)/d\nu$, where $k = n(\nu) \nu/c$. For $\bar N d^{2}/\gamma\hbar \sim 0.1$, i.e. $\bar N \sim 10^{12}cm^{-3}$, $\nu/\gamma \sim 10^{8},$ $n_{g}$ may reach values of  order of $10^{7}$ of either sign [see Figs.~({\ref{fig-3ab}b,d}) where $\chi^{''}=0$]. The refractive index may take values below unity at such moderate atomic densities. Thus, by properly choosing the external parameters one can arrive at rather low subluminal or large superluminal group velocities. 

In summary we have demonstrated that collective interactions among two-level radiators are suitable to generate highly refractive media with GHz switching times to strongly deviating properties. Further, the group velocity of a weak electromagnetic field pulse probing the laser-driven atomic sample may be abruptly altered depending sensitively on the external atomic and laser parameters.


\section*{References}

\begin{thebibliography}{28}
\bibitem{Moll} B. R. Mollow  1969, Phys. Rev. {\bf 188}, 1969; B. R. Mollow 1972, Phys. Rev. A {\bf 5}, 2217; F.Y.Wu et. al. 1977, Phys. Rev. Lett. {\bf 38}, 1077.
\bibitem{Boyd} A. Suguna and G. S. Agarwal 1979, Phys. Rev. A {\bf 20}, 2022; R. W. Boyd et al. 1981, {\it ibid.} {\bf 24}, 411; R. S. Bennink et. al. 2001, {\it ibid.} {\bf 63}, 033804; M. E. Crenshaw and C. M. Bowden 1991, Phys. Rev. Lett. {\bf 67}, 1226.
\bibitem{H_EIT} S. E. Harris, J. E. Field, and A. Imamoglu 1990, Phys. Rev. Lett. {64}, 1107; S. E. Harris 1997, Phys. Today {\bf 50}(7), 36.
\bibitem{LWI} O. A. Kocharovskaya and Y. I. Khanin  1988, JETP. Lett. {\bf 48}, 630; S. E. Harris  1989, Phys. Rev. Lett. {\bf 62}, 1033; M. O. Scully, S.-Y. Zhu, and A. Gavrielides  1989, {\it ibid.} {\bf 62}, 2813; A. S. Zibrov et. al.  1995, {\it ibid.} {\bf 75}, 1499.
\bibitem{refr1} M. O. Scully  1991, Phys. Rev. Lett. {\bf 67}, 1855; M. Fleischhauer et. al. 1992, Phys. Rev. A {\bf 46}, 1468;  H. Friedmann and A. D. Wilson-Gordon 1993, Opt. Commun. {\bf 98}, 303; O. Kocharovskaya, P. Mandel, and M. O. Scully 1995, Phys. Rev. Lett. {\bf 74}, 2451. 
\bibitem{refr2} T. Quang, H. Freedhoff  1993, Phys. Rev. A {\bf 48}, 3216; C. Szymanowski and C. H. Keitel  1994, J. Phys. B: At. Mol. Opt. Phys. {\bf 27}, 5795; U. Akram, M. R. B. Wahiddin, Z. Ficek  1998, Phys. Lett. A {\bf 238}, 117; M. Haas, C. H. Keitel 2003, Optics Commun. {\bf 216}, 385; G. Li et. al. 2000, J. Phys. B {\bf 33}, 3743.
\bibitem{stop} O. Kocharovskaya, Y. Rostovtsev, M. O. Scully 2001, Phys. Rev. Lett. {\bf 86}, 628; D. F. Philips et. al.  2001, {\it ibid.} {\bf 86}, 783.
\bibitem{ultra} Peng Zhou and S. Swain  1996, Phys. Rev. Lett. {\bf 77}, 3995; C. H. Keitel 1999, {\it ibid.} {\bf 83}, 1307.
\bibitem{HarPl} S. E. Harris 1996, Phys. Rev. Lett. {\bf 77}, 5357.
\bibitem{AgL} G. S. Agarwal, Robert W. Boyd  1999, Phys. Rev. A {\bf 60}, R2681.
\bibitem{Man} A. S. Manka et al. 1994, Phys. Rev. Lett. {\bf 73}, 1789.
\bibitem{mek} M. Macovei, J. Evers, and C. H. Keitel 2003, Phys. Rev. Lett. {\bf 91}, 233601; M. Macovei, J. Evers, and C. H. Keitel 2005, Phys. Rev. A {\bf 71}, 033802.
\bibitem{Dan} A. Wicht, R.-H. Rinkleff, L. Spani Molella, and K. Danzmann  2002, Phys. Rev. A 66, 063815.
\bibitem{Andr} A. V. Andreev, V. I. Emel'yanov, and Yu. A. Il'inskii, {\it Cooperative Effects in Optics. Super\-fluorescence and Phase Transitions} (IOP Publishing, London, 1993).
\bibitem{Puri} R. R. Puri, {\it Mathematical Methods of Quantum Optics} (Springer, Berlin 2001).
\bibitem{Dicke} R. H. Dicke  1954, Phys. Rev. {\bf 93}, 99.
\bibitem{Lexp} H.-A. Bachor and T. C. Ralph, {\it A Guide to Experiments in Quantum Optics} ( WILEY-VCH, Weinheim 2004).
\bibitem{SZb} M. O. Scully and M. S. Zubairy, {\it Quantum Optics}, Cambridge University Press (1997).
\bibitem{ACS} R. Bonifacio, Dae M. Kim and Marlan O. Scully  1969, Phys. Rev. {\bf 187}, 441; R. Gilmore, C. M. Bowden and L. M. Narducci 1975, Phys. Rev. A {\bf 12}, 1019; G. S. Agarwal et. al. 1979, Phys. Rev. Lett. {\bf 42}, 1260.
\end{thebibliography}
\end{document}